\documentstyle[twocolumn,aps,prl]{revtex}

\begin{document}
\draft
\title{ Localization Characteristics of Relativistic vs Nonrelativistic Fermions on a Lattice} 
%of Dirac fermions  with disordered Mass }
\author{Indubala I. Satija\cite{email}}
\address{
 Department of Physics, George Mason University,
 Fairfax, VA 22030}
\date{\today}
\maketitle
\begin{abstract}
Relativistic fermions on a lattice are shown to correspond to the
$fluctuations$ in the localized nonrelativistic fermions.
Therefore, in contrast to nonrelativistic case,
the relativistic fermions are critical with universal exponents
described by the strong coupling limit of the nonrelativistic problem.
The fluctuations 
also describe $anisotropic$ spin chain at the onset to {\it long range magnetic order}
whose universality class is the Ising model.
This generalizes the universality in spin models to include multifractal exponents.
Finally, analogous to the nonrelativistic case, 
the relativistic fermions may exhibit ballistic character due to correlations. 

\end{abstract}

\narrowtext

The problem of Dirac fermions with random mass has been the subject of
various theoretical studies.\cite{fisher,steiner,gogolin} 
For weak disorder, the model describes one-dimensional (1D) spinless fermions with
random hopping matrix elements\cite{fisher}, a toy model  
that exhibits localization transition in 1D.
Both the spin-Peierls and the spin-ladder systems with non-magnetic impurities
have been described by Dirac equation with random mass.\cite{steiner,gogolin}
The zero-energy states which dominate the low temperature thermodynamic have been the subject of main focus
as it provides an exactly solvable model whose multifractal spectrum can be computed analytically
.\cite{Fradkin,Alain}
It turns out that in contrast to the usual 1D localization problem for nonrelativistic fermions, correlation functions
for relativistic fermions exhibit power-law decay with universal exponents.\cite{steiner}

The central result of this paper is that
the differences between the 
relativistic and the nonrelativistic fermions in presence of disorder, can be attributed to simple relationships
between these two problems.
We show that relativistic fermions are described by 
the {\it fluctuations in the exponential decaying wave functions of the nonrelativistic fermions}.
New insight into both
the absence of localization and the universality of zero mode Dirac fermion is gained
by their mapping to the anisotropic spin-$1/2$ chain at the onset to long range order(LRO).
Zero mode Dirac fermions describe critical spin chains with $O(1)$ symmetry 
while the corresponding non-relativistic fermions are related to $O(2)$ spin chains.
We argue that the massless excitations of the spin chain responsible for
long range magnetic correlations provide mechanism for delocalization of relativistic fermions and the statement
of universality is the well known statement that the anisotropic spin chains belong to
the universality class of the Ising model.
Although the setting we describe is quite general in the context of disordered systems,
for concreteness we will consider
quasiperiodic disorder which exhibits localization transition in nonrelativistic lattice problem.
The reason for this particular choice is that the
results from exact renormalization group (RG)\cite{KSprl} as well as rigorous mathematical analysis\cite{andy} can be
used to compare and contrast the role of disorder in relativistic vs nonrelativistic problem.
For quasiperiodic disorder, non-relativistic problem known as Harper equation\cite{Harper} 
has been studied extensively. 
Another interesting result of this paper is that
it provides a new way to understand recently discussed strong coupling fixed point of Harper equation
which describes the universal
aspects of the localized phase.\cite{KSprl} The strong coupling universality 
is the statement that the universality class of anisotropic spin chain is determined by the Ising limit.

We consider a nearest-neighbor (nn) tight binding model (tbm) of spinless non-interacting
fermions in a disordered
potential $V_n$,
\begin{equation}
H=\sum [ c_{n+1}^{\dag} c_n+2\lambda V_n c^{\dag}_n c_n ]
\end{equation}
Here the $c$'s are canonical Fermion operators and
$\lambda$ is the strength of the aperiodic potential $V(x)$. The 
eigenstates of this lattice system are given by the following discrete Schrodinger equation,
\begin{equation}
\psi_{n-1}+\psi_{n+1}+2 \lambda V_n \psi_n = E \psi_n.
\end{equation}

For disordered systems exhibiting exponential localization,
the fermion wave functions $\psi_n$ can be written as{\cite{KSprl}
\begin{equation}
\psi_n = e^{ -\gamma |n|} \eta_n
\end{equation}
where $\gamma$ is the inverse localization length $\gamma=\xi^{-1}$. 
The tbm describing the fluctuations $\eta_n$ in the exponentially decaying
envelope is given by the following pair of equations,

\begin{eqnarray}
e^{-\gamma} \eta^r_{n+1} + e^{\gamma} \eta^r_{n-1} + 2 \lambda V_n
\eta^r_n &=& E \eta^r_n \nonumber\\
e^{\gamma} \eta^l_{n+1} + e^{-\gamma} \eta^l_{n-1} + 2 \lambda V_n
\eta^l_n &= &E \eta^l_n
\end{eqnarray}

Here $\eta^r$ and $\eta^l$ respectively describe the fluctuations 
to the right and to the left of the localization
center, which is chosen to be at $n=0$. As shown below, these equations describe relativistic fermions on a lattice 
for zero energy states
as in the long wave length limit, the equations reduce to the Dirac equation. 
We replace $n$ by $x$ and write 
$\eta_{n\pm 1} = e^{\pm ip} \eta(x)$, where $p$ is the momentum canonically conjugate to $x$.
The equation (4) for the fluctuations for $E=0$ state can be described by the following non-Hermitian Hamiltonian $H_{fluc}$
and its adjoint,\cite{footnote}

\begin{equation}
H_{fluc} = e^{-\gamma} e^{i p} + e^{\gamma} e^{-i p} + 2\lambda V(x)
\end{equation}
%Therefore,  $\eta^r$ and $\eta^l$ are complex conjugate of each other.
In the limit ($p \rightarrow 0$), 
the system for $E=0$ reduces to the Dirac equation,
\begin{equation}
[g \sigma_x p -i(2\lambda V(x)+2 cosh(\gamma) )\sigma_y]\eta(x) = 0
\end{equation}
where $\eta(x)$ is a two-dimensional spinor
$\eta(x)=(\eta^l(x),\eta^r(x))$. 
Therefore, the two-component structure of Dirac spinor arises naturally when we consider fluctuations about exponentially
localized wave functions.
Here $g \equiv 2sinh(\gamma)$ is the velocity of the Dirac fermions while the mass of the Dirac fermions
$m(x) = 2((\lambda V(x)+ cosh(\gamma))$. The $\sigma_k$, $k=x,y$ are the Pauli matrices.
%Hence, $H_{fluc}$ is the Dirac Hamiltonian on a lattice for $E=0$ states.

Therefore, on a lattice, the Dirac fermions 
with disordered mass are the fluctuations of the nonrelativistic localized fermions. 
This would imply the absence of exponential localization for relativistic fermions provided
the equation (4) has a solution with $E=0$.

Before we discuss the solution of the discrete Dirac equation, 
we show the equivalence between the disordered Dirac Hamiltonian 
and a spin chain with disordered magnetic field. The anisotropic XY spin-$1/2$ chain in a transverse magnetic field is,
\begin{equation} \label{spin}
H=-\sum [e^{-\alpha} \sigma^x_n \sigma^x_{n+1}+ e^{\alpha}\sigma^y_n \sigma^y_{n+1} + 2\lambda V_n \sigma^z_n]. 
\end{equation}
The $e^{-\alpha}$ and $e^{\alpha}$ respectively describe the exchange interactions along the
$x$ and the $y$ directions in spin space and
therefore, the spin space anisotropy $g$ is given by $g=2sinh(\alpha)$.
It is well known that the Jordan-Wigner transformation, 
\cite{Lieb} transforms the spin problem to spinless fermion problem where fermions are the quasiparticle
excitations of the spin chain. The eigenstates of the excitations
are described by the following coupled equations,\cite{Lieb,KSrg}

\begin{eqnarray}\label{fermion}
e^{-\alpha} \eta^1_{n+1}+e^{\alpha}  \eta^1_{n-1}+2\lambda V_n \eta^1_i &=& E \eta^2_n \nonumber\\
e^{\alpha}  \eta^2_{n+1}+e^{-\alpha}  \eta^2_{n-1}+2\lambda V_n \eta^2_n &=& E \eta^1_n
\end{eqnarray}

For $E=0$, these equations reduce to equation (4) with the parameter $\alpha$ equals the
inverse localization length $\gamma$. However, $E=0$ is the solution of this equation
only at the onset to LRO, and therefore, for anisotropic spin chain with disordered field $V_n$,
this mapping between the spin and fermions provides an interesting way to describe 
the onset to LRO: the massless excitations of the critical anisotropic spin chain describes the
fluctuations in the exponentially localized excitations of the isotropic chain provided the anisotropy
parameter is the inverse localization length.
Alternatively, the mapping provides a new method to determine the localization length of
tbms in the presence of disorder. 
For the case of quasiperiodic disorder,
$V_n = cos(2 \pi (\sigma n+\phi))$, ( where $\sigma$ is an irrational number and $\phi$
is a constant phase factor), the localization length is
$\xi^{-1}=\log(\lambda)$.\cite{Harper}
Therefore $\alpha=log(\lambda)$
is the equation for the critical line, a result known from earlier numerical study\cite{IIS} for the spin chain.

The Ising limit ($\alpha \rightarrow \infty$) of the critical spin chain coincides with the
$\lambda \rightarrow \infty$ limit of the Harper equation.   
This is the strong coupling limit of the Harper equation\cite{KSprl}
that has been studied by
exact RG as well
as more rigorous analysis\cite{andy}. 
The existence of a  strong coupling
fixed point shows the universality of the fluctuations in the exponentially decaying solutions
in the localized phase. The equivalence between the strong coupling limit of Harper equation and the
critical Ising spin in a quasiperiodic transverse field
provides a new way to understand this fixed point:
the statement of universality of the strong coupling
fixed point of Harper equation is equivalent to the statement that the anisotropic spin chain in a quasiperiodic
field  belongs to the universality class
of the Ising model, a result well known for periodic spin systems in arbitrary dimension.

We now focus on the $E=0$ solution of the Harper equation in the localized phase as the fluctuations
about this describe the Dirac fermions. The universal features of this solution
can be studied by an exact RG scheme.\cite{KSrg}
The earlier studies of this problem was confined to the band edges of the eigenspectum.
To show the universal features of the solution, with $\sigma $ equals inverse golden mean,
all sites except those labeled by the Fibonacci
numbers $F_m$ are decimated.
At the $m^{th}$ decimation level,
the tbm describing the fluctuations is expressed in the form
\begin{equation}
f_m(n) \eta_{n+F_{m+1}}=\eta_{n+F_m} + e_m(n) \eta_n.
\end{equation}
The additive property of the Fibonacci numbers provides an exact recursion
relations for the decimation functions $e_m$ and $f_m$.
In the strong coupling limit $f_m$ is found to approach zero 
simplifying the RG flow as,\cite{KSprl}
\begin{equation}
e_{m+1}(n) = - e_{m-1}(n+F_m) e_m(n)
\end{equation}
Numerical iteration of this equation provides an extremely accurate method to distinguish extended, 
localized and critical states.
The extended and localized phase
correspond to trivial asymptotic RG flow while
the critical behavior is found to be
characterized by a nontrivial
asymptotic $6$-cycle in the decimation function: $e_{m+6}=e_m$. 

Figure 1 shows the self-similar fluctuations about exponentially localized solutions in Harper equation,
which describes zero-mode Dirac problem with quasiperiodic mass. 
In the strong coupling limit, 
which also describes the massless mode of the Ising chain
the wave function is given by the following algebraic equation,
\begin{equation}
\frac{\eta_n} {\eta_0} \equiv exp-[\Phi(n)] = exp-[\sum_{j=1}^n log(2|V_j|)]
\end{equation}
For random disorder, the above summation can be done explicitly resulting in exact solution
\cite{Fradkin,Alain}, that
facilitated an exact calculation of multi-fractal spectrum.

In order to seek the physical meaning of universal $3$-cycle, and see 
whether the statement that the critical exponents of spin chains are determined by the symmetry of the spin chain
applies to the multifractal exponents,
we compute the $f(\alpha)$ curve describing the multifractal spectrum
associated with the self-similar wave function or
the inverse participation ratios ${\bf P}$, 
\begin{eqnarray}
{\bf P}(q,N) & = &\frac{\sum |\eta_n|^{2q}}{\sum |\eta_n|^2} \sim N^{-\tau(q)}\nonumber\\
\alpha &=& \frac{d\tau}{dq}\nonumber\\
f(\alpha) &=& \alpha q-\tau(q)
\end{eqnarray}
The free energy function $\tau(q)$ and
its Legendre transform $f(\alpha)$ were found to be $\lambda$ independent  only for {\it for positive values of $q$}
and hence only left half of the $f(\alpha)$ curve is universal.
Therefore, for quasiperiodic spin chains at the onset to LRO, scaling exponents for {\it only
positive moments} of the participation ratio are
determined by the spin space symmetry.

We next show that the presence of correlated disorder can result in propagating solutions for the disordered
Dirac equation, as is known to be the case for the non-relativistic fermions.\cite{dimer}
For quasiperiodic disorder, dimer-type correlations can be introduced by
replacing $\theta_n=2\pi\sigma n$ in $V(\theta_n)$ by the iterates of the $supercritical$
standard map, describing 
Hamiltonian systems with two degrees of freedom.\cite{mackay,dimerlong}
\begin{equation}
\theta_{n+1}+\theta_{n-1}-2\theta_n = - {K\over  2\pi } \sin( 2 \pi \theta_n) .\label{SM}
\end{equation}
We use iterates that describe golden-mean cantous, 
which has been shown to exhibit dimer-type correlations, and leads to
Bloch-type states for the nonrelativistic fermions.\cite{dimerlong}
Here, we will confine to the Ising model, described by,
\begin{eqnarray}
\eta^1_{n+1}+2\lambda cos(2\pi\theta_n)\eta^1_n&=& E \eta^2_n\nonumber\\
\eta^2_{n-1}+2\lambda cos(2\pi\theta_n)\eta^2_n&=& E \eta^1_n
\end{eqnarray}
We determine the critical $\lambda$ as a function of $K$, and analyze the massless mode of the Ising model using the 
RG methodology\cite{KSrg}. As shown in figure 3, nontrivial $6$-cycle degenerates to trivial fixed points
at certain special parameter values.
The origin of these trivial fixed points, has been traced to
a {\it hidden dimer} in the quasiperiodic iterates describing the golden-cantorus.\cite{dimerlong}
At these points, the relativistic massless mode of the Ising model is ballistic.
Further details of the effects of correlated disorder on magnetic and spectral transitions will
be discussed elsewhere.

We would like to point out that although the relationship between the fermion and the spin problem is well
known, what is new here is the that disordered spin chains
provide important key in understanding the role
of disorder in transport properties of relativistic and non-relativistic fermions.
The differences between these two fermions is traced to the differences between spin chains with $O(1)$
and $O(2)$ symmetry where the spin chains with $O(1)$ symmetry exhibits magnetic long range correlations.
The correspondence with the spin chain also explains why $E=0$ mode of Dirac equation is special:
the root of criticality of $E=0$ Dirac fermions can be attributed to the fact that
it corresponds to infinite magnetic correlation length.
We argue that these magnetic correlations account for the absence of exponential localizations
in relativistic fermions.
It is rather interesting that
the relativistic fermions can be viewed as the fluctuations
about localized nonrelativistic fermions and this picture also explains the absence of localization of Dirac fermions.
Furthermore, the universal aspects of relativistic fermions are described by the strong coupling limit
of the nonrelativistic fermions. This universality implying that the positive moments of the participation ratio
have universal exponents , generalizes the well known statements of universality of periodic spin chains
to quasiperiodic spins exhibiting multifractal exponents.
Finally, we would like to mention that many of our conclusions are valid for random as well
as pseudorandom disorder.\cite{footSS}

This research is supported by a grant from National Science Foundation,
DMR-0072813.
I would like to thank Alain Comtet for various discussions and bringing to my attention
the problem of Dirac fermions with random mass.

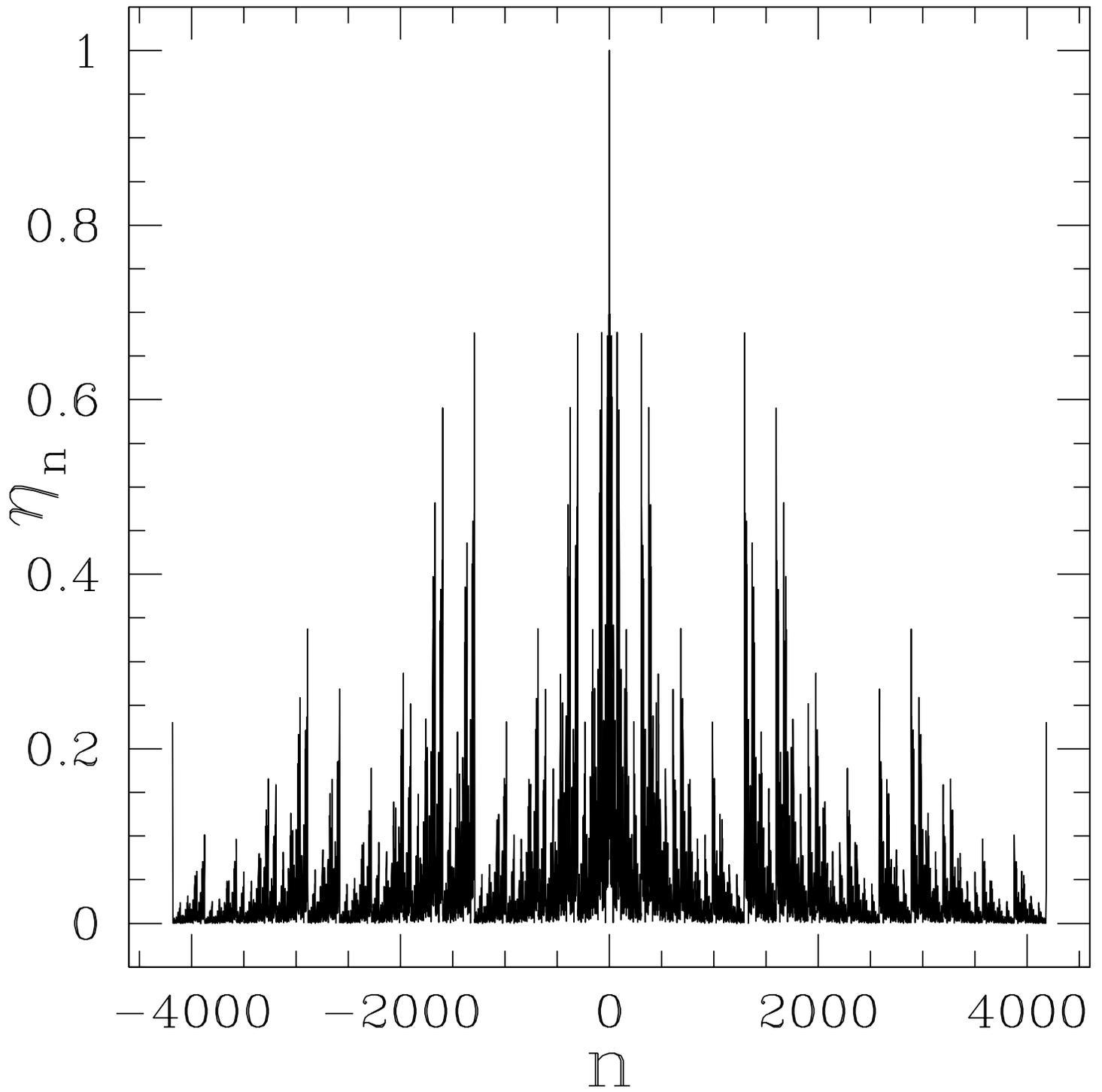
\begin{figure}
\caption{ (a) Absolute value of the fluctuations for Harper equation for $E=0$ states with $\phi=.25$
. At the Fibonacci sites,
we see period-$6$ behavior(period-$3$ in absolute value): $|\eta_{F_m}|=|\eta_{F_{m+3}}|$ which 
is independent of $\lambda$. } 
\label{fig1}
\end{figure}

\begin{figure}
\caption{Numerically obtained (a) $f(\alpha)$ curves for $\lambda \rightarrow \infty$( solid curve) and 
$\lambda=1.5$ (lines with crosses).} 
\label{fig2}
\end{figure}

\begin{figure}
\caption{(a)Critical $\lambda$ as a function of $K$ for the Ising model.
(b) RG $6$-cycle showing the variation in the renormalized coupling $e$ after Fibonacci decimation.}
\label{fig3}
\end{figure}


\begin{references}

\bibitem{email} e-mail: isatija@gmu.edu.

\bibitem{fisher} L. Balents and M. Fisher, Phys Rev B, 56, 12970 (1997).

\bibitem{steiner} M. Steiner, M. Fabrizio and A. O Gogolin, Phys Rev B, 57 (1998) 8290.
M. Steiner, Y Chen, M. Fabrizio and A. O Gogolin, Phys Rev B 59(1999) 14848.

\bibitem{gogolin} A. O. Gogolin, A. O. Nersesyan, A. M Tsvelik and L Yu, cond-mat/9707341

\bibitem{Fradkin} H. Castillo, C. Chamon, E. Fradkin, P. Goldbart and C. Mudry, Phys Rev B, 56, (1997) 10668.

\bibitem{Alain}A. Comtet, S. Nechaev and R. Voituriez, cond-mat/0004491, 2000

\bibitem{KSprl} J. A. Ketoja and I. I. Satija, Phys. Rev. Lett. 75,
2762 (1995).

\bibitem{KSrg}
 J. A. Ketoja and I. I. Satija, Phys. Lett. A 194, 64
(1994), Physica A, 219, 212, (1995).

\bibitem{andy}B. D. Mestel, A. H. Osbaldestin and B. Winn;" Golden mean renormalization
for the Harper equation: the
strong coupling fixed point", preprint 2000.
 
\bibitem{Harper} P. G. Harper, Proc. Phys. Soc. London A 68, 874 (1955);
B. Simon,
Adv. Appl. Math. 3, 463 (1982);
J.B. Sokoloff, Phys. Rep. 126, 189 (1985).

\bibitem{IIS} I.I. Satija, Phys Rev B49, 3391 (1994).

\bibitem{footnote} The non-Hermitian Hamiltonian (5) 
can be viewed as a disordered system with complex vector potential that was studied recently.
(N. Hatano and D. R Nelson, Phys. Rev. Lett.
77, 570 (1996); Phys Rev B 56, 8651 (1997).)
It is rather interesting to note that
if $\gamma$ is the inverse localization length, this non-Hermitian Hamiltonian has real spectrum.

\bibitem{dimer}H-L. Wu, W. Golf and P. W. Phillips, Phys Rev B 45 (1992) 1623.

\bibitem{dimerlong} J. Ketoja and I. Satija, Phys Rev B, 59, 9174 (1999);

\bibitem{mackay}R Mackay, Ph.D thesis ( 1983). Princeton Univ.

\bibitem{Lieb}E. Lieb, T. Schultz, and D. Mattis, Ann Phys. (N.Y) 16,
407 (1961).

\bibitem{footSS} The
fractal character of fluctuations has also been discussed in pseudorandom case
with Lloyd distribution: I. Satija, B. Sundaram and J. Ketoja,
, Phys Rev E, 60,(1999) 453.
\end{references}
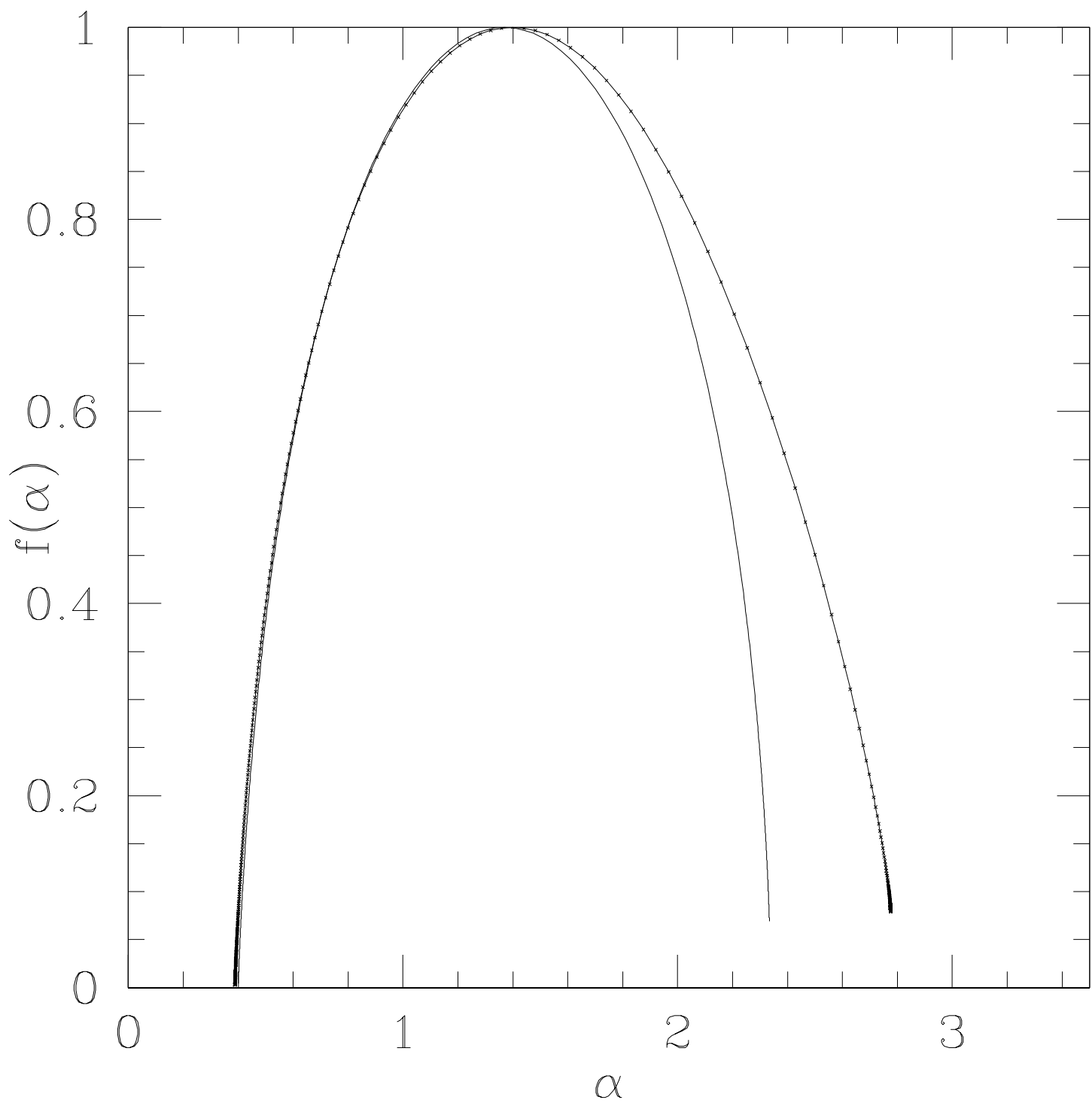
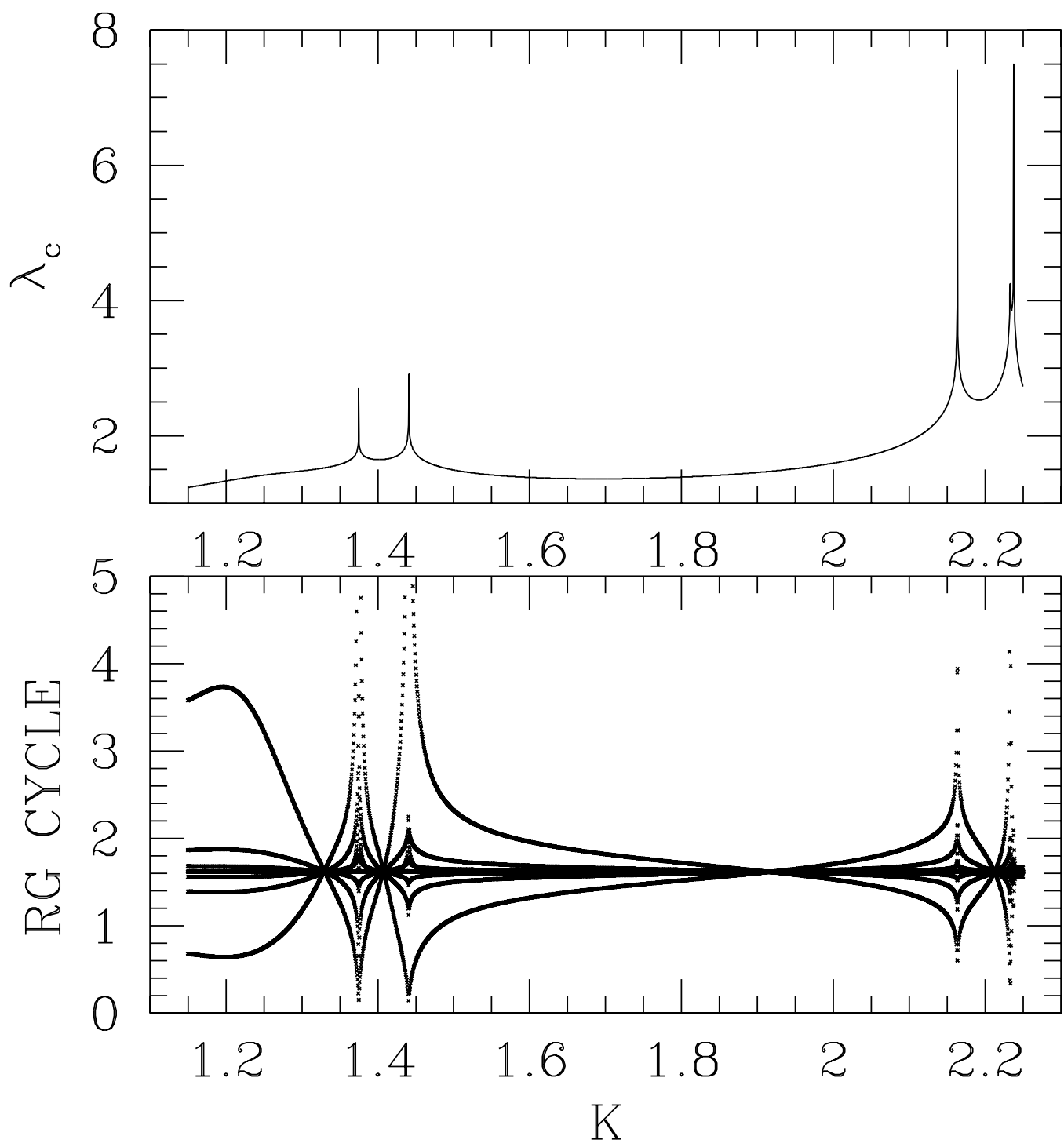
\end{document}